\documentstyle[aps,epsfig]{revtex}   
\newcommand{\vp}{{\vec p}}
\begin{document}

\preprint{Preprint Number: \parbox[t]{45mm}{ANL-PHY-9434-TH-99}}

\title{Memory effects and thermodynamics in strong field plasmas}
\author{J.C.R. Bloch, C.D. Roberts and S.M. Schmidt\vspace*{0.2\baselineskip}}
\address{Physics Division, Argonne National Laboratory,\\ Argonne, Illinois
60439-4843, USA\\[0.6\baselineskip]
\parbox{140mm}{\rm We study the evolution of a strong field plasma using a
quantum Vlasov equation with a non-Markovian source term and a simple
collision term, and calculate the time dependence of the energy- and
number-density, and the temperature.  The evolution of a plasma produced with
RHIC-like initial conditions is well described by a low density approximation
to the source term.  However, non-Markovian aspects should be retained to
obtain an accurate description of the early stages of an LHC-like
plasma.\\[0.4\baselineskip] Preprint Number: ANL-PHY-9434-TH-99,
Pacs Numbers: 12.38.Mh, 25.75.Dw, 05.20.Dd, 05.60.Gg }}

\twocolumn
%
%
%
\maketitle
%


At extreme temperature, hadronic matter undergoes a phase transition to an
equilibrated quark gluon plasma.  Estimates from phenomenological models and
lattice-QCD indicate a critical temperature for this transition: $T_c \sim
170\,$MeV, which corresponds to an energy density of $2$-$3\,$GeV/fm$^3$.  It
is hoped that this plasma will be produced at the Relativistic Heavy-Ion
Collider (RHIC) and/or the Large Hadron Collider (LHC)\cite{mum}.
However, little is currently understood about the formation, equilibration
and hadronisation of the plasma, and herein we focus on dynamical aspects of
the pre-equilibrium phase.

We employ a flux-tube model to explore the creation of a strong-field plasma
and follow its evolution towards equilibrium.  The model assumes that the
collision of two heavy nuclei produces a strong background field and a region
of high energy density, which decays via pair emission.  The particles
produced in this process are accelerated by the background field, providing a
current and a field that opposes the background field.  This is the
back-reaction process, which may result in plasma oscillations.  For QED in
an external field it has been studied via mean field methods and using a
quantum Vlasov equation with a Schwinger-like source term\cite{Back}.
Collisions between the particles damp the plasma oscillations and are
necessary to equilibrate the plasma, and their effect has been modelled in
the Vlasov equation approach\cite{KM,bahl,eis,nayak}.

For strong fields, non-Markovian aspects of the particle production mechanism
are very important\cite{Rau,kme,basti,prd}.  Herein we emphasise this, using
a relativistic transport equation with the non-Markovian source term derived
in Refs.~\cite{gsi,kme,basti} and explored in Refs.~\cite{prd,bloch}.  We
apply it with impact energy densities of the scales anticipated at RHIC and
LHC, and study the time evolution of the plasma's properties.

We model the effect of the nucleus-nucleus collision by an external,
spatially-homogeneous, time-dependent Coulomb-gauge vector potential, which
defines the longitudinal direction: $A_\mu= (0,0,0,A(t))$.  
The kinetic equation describing fermion production in this external field
is\cite{gsi} 
\begin{equation}
\label{10}
\frac{df(\vp,t)}{dt}=S(\vp,t) + C(\vp,t) 
\end{equation}
where, in contrast to the Schwinger production rate, the source term here is
momentum- and time-dependent:
\begin{eqnarray}\label{source}\nonumber
&&S(\vp,t) = \frac{1}{2}
\frac{eE(t)\varepsilon_\perp}{\omega^2(t)}
\int_{-\infty}^t dt'\frac{eE(t')\varepsilon_\perp}{\omega^2(t')}\\
&&\times [1-2f(\vp,t')] \cos[2(\Theta(t)-\Theta(t'))]\,,
\end{eqnarray}
$E(t) = -dA(t)/dt$ and $e$ is an electric charge, with the dynamical phase
and total energy, respectively,
\begin{eqnarray}
\Theta(t)  & = & \int^t_{-\infty}dt'\omega(t')\,,\\
\omega(t) & = & \sqrt{\varepsilon_\perp^2+(p_\parallel-eA(t))^2}\,,
\end{eqnarray}
where $\varepsilon_\perp=\sqrt{m^2+\vp_\perp^{\,2}}$ is the transverse
energy.  $m^2\sim \Lambda_{QCD}^2\sim 0.2$ GeV/fm sets a typical scale.

The second term on the right-hand-side (r.h.s.) of Eq.~(\ref{10}) describes
collisions between the particles, and we employ a simple and widely-used
model
\begin{equation}
\label{collterm}
C(\vp,t) =\frac{f^{eq}(\vp,t) - f(\vp,t)}{\tau_r}\,,
\end{equation}
where $\tau_r$ is the ``relaxation time'' and $f^{eq}$ is the thermal
equilibrium distribution functions for fermions:
\begin{equation}
\label{feq} f^{eq}(\vp,t) =
\frac{1}{\exp[\omega(\vp,t)/T(t)]+ 1}\,. 
\end{equation}
The temperature profile in Eq.~(\ref{feq}) is {\it a priori} unknown and is
determined by requiring that the average energy in our ensemble is that of a
quasi-equilibrium fermion gas, Eqs.~(\ref{Tprof})-(\ref{neq}).  We note that
the relaxation time approximation is only valid under conditions of local
``quasi-equilibrium''.  That is difficult to justify in the presence of
strong fields.  Further, a realistic collision term is likely to generate
memory effects additional to those present in the source term, which are our
present focus.  The improvement of the collision term is therefore an
important contemporary challenge.

The kinetic equation, Eq.~(\ref{10}), is non-Markovian for two reasons: (i)
the source term on the r.h.s. requires knowledge of the entire history of the
evolution of the distribution function from $t_{-\infty}\rightarrow t$; and
(ii), even in the low density limit ($f(t)\approx 0$), the integrand is a
non-local function of time as is apparent in the coherent phase oscillation
term: $\cos[\Theta(t)-\Theta(t')]$.  Our kinetic formulation makes possible a
simple and direct connection with widely used approximations.

In the low density limit the source term is independent of the distribution
function
\begin{eqnarray}
\label{lds}\nonumber
S^0(\vp,t) &=& \frac{1}{2}
\frac{eE(t)\varepsilon_\perp}{\omega^2(t)}\\
&&
\times\int_{-\infty}^t
dt'\frac{eE(t')\varepsilon_\perp}{\omega^2(t')}\cos[2(\Theta(t)-\Theta(t'))]\,.
\end{eqnarray} 
The kinetic equation, Eq.~(\ref{10}), with the source term of Eq.~(\ref{lds})
has the general solution
\begin{eqnarray}
&&f^0_\pm(\vp,t)=\\\nonumber
&&\int_{-\infty}^{\,t}\,dt'\exp\bigg[\frac{t^\prime-t}{\tau_r}\bigg]\bigg(
S_\pm^0(\vp,t')+\frac{f_\pm^{eq}(\vec{p},t^\prime)}{\tau_r}\bigg)\,.
\end{eqnarray}

The effect of back-reactions on the induced field is accounted for by solving
Maxwell's equation: $\dot E(t)=-j(t)$.  The total electric field, $E(t)$, is
the sum of an external field, $E_{ex}(t)$, and an internal field,
$E_{in}(t)$.  Herein we assume that the plasma is initially produced by the
external field, excited by an external current, $j_{ex}(t)$, such as might
represent a heavy ion collision
\begin{equation}
E_{ex}(t)=-A_0[b\cosh^{2}(t/b)]^{-1}\,.
\end{equation}
This model electric field ``switches-on'' at $t\sim -2 b$ and off at $t\sim 2
b$, with a maximum magnitude of $A_0/b$ at $t=0$.  Once this field has
vanished only the induced internal field remains to create particles and
affect their motion.

Continued spontaneous production of charged particle pairs creates a
polarisation current, $j_{pol}(t)$, that depends on the particle production
rate, $S(\vp,t)$.  Meanwhile the motion of the existing particles in the
plasma generates a conduction current, $j_{cond}(t)$, that depends on their
momentum distribution, $f(\vp,t)$.  The internal current is the sum of these
two contributions
\begin{equation}\label{1.14}
{\dot E}_{in}(t)=-j_{in}=-j_{cond}(t)-j_{pol}(t)\,.
\end{equation}

The renormalised Maxwell equation is\cite{bloch}
\begin{eqnarray}\label{dotE}
&&\dot E_{in}(t) = -2e\int
\frac{d^3\vp}{(2\pi)^3}\frac{p_\parallel-eA(t)}{\omega(\vp,t)}\bigg[f(\vp,t)\\
\nonumber
&&+\frac{1}{2}\frac{\omega(\vp,t)}{{\dot
\omega(\vp,t)}}\frac{df(\vp,t)}{dt}
-{\dot E}(t) \frac{\varepsilon^2_\perp}{8\omega^4(p_\parallel-eA(t))}\bigg]\,.
\end{eqnarray}
The first two terms on the r.h.s. represent the conduction and polarisation
current, respectively, while the last term arises in regularising the
polarisation current.  The solution of the coupled pair, Eqs.~(\ref{10}) and
(\ref{dotE}), yields $E(t)$ and $f(\vec{p},t)$, and makes evident effects
such as plasma oscillations and collisional damping\cite{bloch}.

\begin{figure}[t]
\psfig{figure=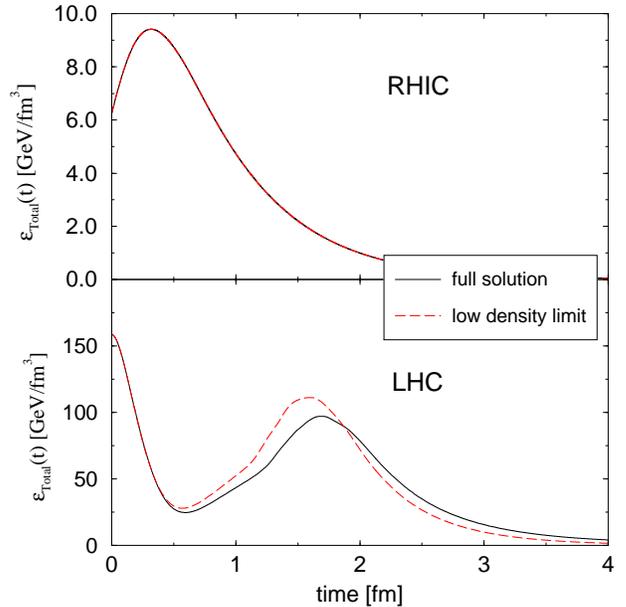,height=8cm,angle=-90}

\vspace{0.5cm}

\caption{Time evolution of the energy density for RHIC (upper panel) and LHC
(lower panel) conditions.\label{fig1} }
\end{figure}
All bulk thermodynamical properties are expressed in terms of $f(\vp,t)$, and
of particular interest herein are: the energy density
\begin{eqnarray}
\epsilon_{\rm Total}(t) & = & \frac{1}{2}E^2(t) + \epsilon(t)\,,\\
\epsilon(t) & = & 2\int \frac{d^3
\vp}{(2\pi)^3}\omega(\vp,t)f(\vp,t)\,, 
\end{eqnarray}
with the pressure $p(t)= 1/3\,\epsilon(t)$; the particle number density
\begin{equation}
n(t) = 2\int \frac{d^3 \vp}{(2\pi)^3}f(\vp,t)\,;
\end{equation}
and the temperature profile: $T(t)$, determined from the condition
\begin{eqnarray}
\label{Tprof}
\frac{\epsilon(t)}{n(t)} & = & \frac{\epsilon^{eq}(t)}{n^{eq}(t)}\,,\\
\epsilon^{eq}(t) & = & 
\int\frac{d^3 \vp}{(2\pi)^3}\,\omega(\vp,t)f^{eq}(\vp,t)\,,\\
\label{neq}
n^{eq}(t) & = & \int \frac{d^3 \vp}{(2\pi)^3}\,f^{eq}(\vp,t)\,,
\end{eqnarray}
which implements our constraint that the average energy in the ensemble is
equal to that of a quasi-equilibrium fermion gas.

To solve the coupled equations, we assume a trial form for $E(t)$, $T(t)$ and
solve for $f(\vec{p},t)$ from Eq.~(\ref{10}).  $f(\vec{p},t)$ so obtained
yields an iterated $E(t)$, $T(t)$ from Eqs.~(\ref{dotE}), (\ref{Tprof}).  The
procedure is repeated until seed and iterate agree within $0.1$\%.

\begin{figure}[t]
\psfig{figure=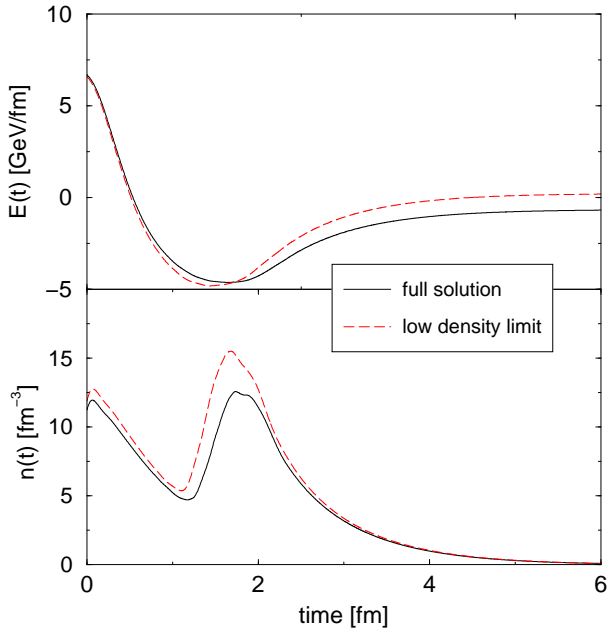,height=8cm,angle=-90}

\vspace{0.5cm}

\caption{Time evolution of the electric field (upper panel) and the particle
density (low panel) for LHC conditions.\label{fig2} }
\end{figure}

Our primary goal is to demonstrate that non-Markovian effects influence the
early stage of plasma evolution.  Therefore we contrast the results obtained
in both the low-density limit, Eq.~(\ref{lds}), and with the full source
term, Eq.~(\ref{source}).  The parameters characterising the external field:
$b=0.5/\Lambda_{\rm QCD} \approx 0.5\,$fm, $A^{RHIC}_0 = 4\,\Lambda_{\rm
QCD}$ and $A_0^{LHC} = 20\,\Lambda_{\rm QCD}$, are chosen in order to obtain
initial field energy densities of the magnitude expected in experiments,
Fig.\ref{fig1}.  We choose a relaxation time $\tau_r = 1/\Lambda_{\rm QCD}
\approx 1\,$fm.

In Fig.~\ref{fig1} we see that for RHIC conditions the energy density rises
rapidly but, after reaching a maximum, decays monotonically.  In this case
the full solution and the solution obtained in the low density limit are
quantitatively identical.  For LHC conditions, with an initial energy density
twenty-times larger, the situation is different.  The solution obtained in
the low density limit is only a qualitative guide to the plasma's behaviour
and plasma oscillations are evident on observable time-scales.  The deviation
between the two curves begins when the strength of the external field wanes.

In Fig.~\ref{fig2} we present the time evolution of the electric field and
particle number density for LHC conditions.  The plasma oscillations are
damped on a time scalar characteristic of the collision time.  The evolution
obtained in the low density limit again provides only a qualitative guide to
the behaviour of the complete solution.  The plasma oscillation frequency and
amplitude are larger in the low density limit because damping is less
effective in the absence of the Pauli blocking factor: $1-2f(\vec{p},t)$.
This same effect is responsible for the overestimate of $n(t)$ in the low
density limit.

\begin{figure}[t]
\psfig{figure=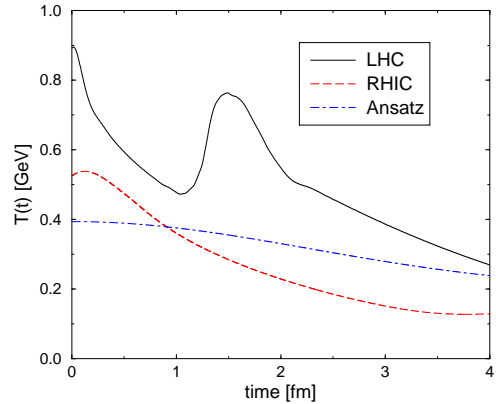,height=6.4cm,angle=-90}
\caption{Time evolution of the quasi-equilibrium temperature for RHIC and LHC
initial conditions.  For comparison we also plot the {\it Ansatz} in
Eq.~(\protect\ref{TAnsatz}).
\label{fig3} }
\end{figure}

Our calculations also yield temperature profiles, which are depicted in
Fig.~\ref{fig3}.  The RHIC-like source conditions yield an initial
temperature $T^{\rm RHIC}(t=0) \sim 0.5\,$GeV, and the temperature decreases
monotonically with increasing $t$.  The LHC-like source conditions yield an
initial temperature twice as large: $T^{\rm LHC}(t=0) \sim 0.9\,$GeV, and the
temperature oscillates in tune with the energy density.  In Ref.~\cite{bloch}
the temperature profile was not determined self-consistently, instead an {\it
Ansatz} was used:
\begin{equation}
\label{TAnsatz}
T(t) = T_{eq} + (T_m - T_{eq})\, {\rm e}^{-t^2/t_0^2}\,,
\end{equation}
with $T_{eq}=\Lambda_{\rm QCD}$, $T_m=2\,T_{eq}$, $t_0^2=10/T_{eq}^2$, which
is also plotted in Fig.~\ref{fig3}.  It is evident that the {\it Ansatz}
provides a not unreasonable model of RHIC-like conditions.

We have solved a quantum Vlasov equation under conditions that qualitatively
mimic those anticipated at RHIC and LHC.  Under RHIC conditions the low
density approximation to the source term provides an accurate description of
the plasma's early stages.  However, the non-Markovian features of the source
term become important when the initial energy density is LHC-like and
generate effects that are likely to be observable.  Following this study we
anticipate that the analogue of our non-Markovian source term would have a
significant effect in a non-Abelian transport equation\cite{nayak}.

\section*{Acknowledgments}
S.M.S. is grateful for the support and hospitality of the Nuclear Theory
Group of the Brookhaven National Laboratory where part of this work was
conducted.  This work was supported by the US Department of Energy, Nuclear
Physics Division, under contract no. W-31-109-ENG-38 and benefited from the
resources of the National Energy Research Scientific Computing Center.
S.M.S. is grateful for financial support from the A.v. Humboldt foundation.


\end{document}